\documentclass[10pt,conference]{IEEEtran}
\usepackage{graphicx} 

\newtheorem{theorem}{Theorem}[section]

\begin{document}

\title{Deriving the Normalized Min-Sum Algorithm from Cooperative Optimization}

\author{\authorblockN{Xiaofei Huang}\\
\authorblockA{School of Information Science and Technology, Tsinghua University, Beijing, P.R. China, 100084 \\
Email: huangxiaofei@ieee.org\\
(Accepted by IEEE Information Theory Workshop, Chengdu, China, 2006)}
}
%

\maketitle

\begin{abstract}
The normalized min-sum algorithm can achieve near-optimal performance at decoding LDPC codes.
However, it is a critical question to understand the mathematical principle underlying the algorithm.
Traditionally, people thought that the normalized min-sum algorithm
	is a good approximation to the sum-product algorithm,
	the best known algorithm for decoding LDPC codes and Turbo codes.
This paper offers an alternative approach to understand the normalized min-sum algorithm.
The algorithm is derived directly from cooperative optimization,
	a newly discovered general method for global/combinatorial optimization.
This approach provides us another theoretical basis for the algorithm and offers new insights 
	on its power and limitation.
It also gives us a general framework for designing new decoding algorithms.
\end{abstract}

\section{Introduction}
LDPC codes can achieve near capacity performance for channel coding.
One popular algorithm for decoding LDPC codes is so called 
	the normalized min-sum algorithm~\cite{Jinghu2002,JinghuChenThesis} (also referred to as the normalized BP-based algorithm).
It is attractive for hardware/software implementations 
	because it reduces the implementation complexity of the sum-product algorithm without losing much of its performance.
The sum-product algorithm~\cite{Kschischang01,Aji97}, 
	is the best known algorithm 
	for decoding both Turbo codes~\cite{Berrou93} and LDPC codes~\cite{Gallager:LDPCC:thesis,MacKay:GCBOVSM}.

The min-sum algorithm~\cite{Wiberg:thesis} in its original form 
	is treated as a generalization of the Viterbi algorithm~\cite{forney73}
	for iterative decoding of code realizations on general graphs.
When a code realization is cycle-free, the algorithm is an exact solution for maximum likelihood decoding.
When there are cycles, it is surprising that the min-sum decoding often works quite well
	in terms of empirical performance.
However, we are lack of theoretical understanding about its remarkable performance in this case.

The sum-product algorithm~\cite{Gallager:LDPCC:thesis} can be viewed 
	as a generalization of the belief propagation  (BP) algorithm developed in AI~\cite{Pearl88}.
The min-sum algorithm is also referred to as the BP-based algorithm~\cite{Fossorier99}
	where the former can be understood as an approximation to the later 
	following the standard approximation of the Max-Log-MAP~\cite{Hagenauer96}.
Compared with the sum-product algorithm,
	the min-sum algorithm sometimes may have a noticeable degradation in performance.
	
The normalized min-sum algorithm was proposed as a better approximation to the sum-product algorithm
	than the original min-sum algorithm.
Simulation results show that the normalized min-sum algorithm
	can improve the performance of the original min-sum algorithm~\cite{Jinghu2002}.
It can also achieve a near-optimal performance in many cases at decoding LDPC codes, 
	very close to the sum-product algorithm.

It might not be straightforward to understand the (normalized) min-sum algorithm 
	if we argue that it is an approximation to the sum-product algorithm.
The sum-product algorithm is only an approximate algorithm 
	for computing the marginal {\it a posteriori} distributions.
The optimal decoding for a channel code, described in Shannon's information theory~\cite{Shannon48},
	is based on finding the codeword of the maximum {\it a posteriori} probability.
The two tasks are closely related to each other, but not equivalent.
Furthermore, we still need more theoretical understanding of the sum-product algorithm 
	even though there are some remarkable progresses recently~\cite{Yedidia05,Wainwright05,Richardson:DOCAILDPCC}.

This paper takes different approach to understand the (normalized) min-sum algorithm.
It will show that the algorithm can be derived as a cooperative optimization algorithm~\cite{HuangBookCCO}.
Cooperative optimization is a newly discovered global/combinatorial optimization method for attacking hard optimization problems.
It breaks a hard optimization problem into a number of sub-problems and solves them together in a cooperative way.
It does not struggle with local minima,
	has global optimality conditions for recognizing global optima,
	and offers us a complete departure from the classic optimization methods.
With proper settings, a cooperative optimization algorithm has a unique equilibrium 
	and converges to it with an exponential rate regardless of initial conditions and perturbations.
	
Deriving the normalized min-sum algorithm as a cooperative optimization algorithm
	can offer us new insights about the algorithm.
Following this approach, this paper attempts to answer a number of important questions related to the min-sum algorithm;
	1) why can it find the optimal codeword?
	2) whether a found codeword is the optimal one?
	3) what is its objective function to be optimized?
A general framework is also presented in this paper for designing new decoding algorithms
	in a systematic way.

\section{Cooperative Optimization}

Cooperative optimization is a general principle 
	for finding the global optimum of a multivariate function $E(x_1, x_2, \ldots, x_n)$.
It utilizes another function in simple forms, such as $\Psi(x_1, x_2, \ldots, x_n) = \sum_i \Psi_i(x_i)$, 
	and iteratively refines the function $\Psi(x)$ as the lower bound of the multivariate function $E(x)$.
At a given iteration $k$, if the lower bound function $\Psi^{(k)}(x)$ has been tightened enough 
	so that its global minimum equals to the global minimum of $E(x)$,
	i.e.,
\[ \min_{x} \Psi^{(k)}(x) = \min_{x} E(x) \ , \]
	then the global minimum of $E(x)$
	is found which is the same as the global minimum of $\Psi^{(k)}(x)$, 
\[ \arg \min_{x} E(x) = \arg \min_{x} \Psi^{(k)}(x) \ . \]
The global minimum of $\Psi^{(k)}(x)$ of the form $\sum_i \Psi^{(k)}_i(x_i)$ can be easily found as
\[ x^{*(k)}_i(\Psi^{(k)}(x)) = \arg \min_{x_i} \Psi^{(k)}_i (x_i), \quad \mbox{for $i = 1, 2, \ldots, n$} \ . \]

Assume that the multivariate function $E(x)$, often referred to as the energy function or the objective function,
	can be decomposed as the aggregation of a number of sub-objective functions,
\[ E(x) = \sum_i E_i(x) \ . \]
Assume further that $\Psi^{(k-1)}(x)$ of the form $\sum_i \Psi^{(k-1)}_i(x_i)$ 
	is a lower bound function of $E(x)$, $\Psi^{(k)}(x) \le E(x)$.
Let $\Psi^{(k)}_i(x_i)$, for $i=1,2,\ldots, n$, be computed as follows
\begin{equation}
\Psi^{(k)}_i (x_i) = \min_{ X_i \setminus{x_i}} (1 - \lambda_k ) E_i (x)  + \lambda_k \sum_{j} w_{ij} \Psi^{(k-1)}_j(x_j) \ ,
\label{cooperative_optimization}
\end{equation}
where $X_i$ the set of variables contained in $E_i(x)$
	and $\min_{ X_i \setminus{x_i}}$ stands for minimizing with respect to all variables in $X_i$ excluding $x_i$.
Then the new function $\Psi^{(k)}(x) = \sum_i \Psi^{(k)}_i(x_i)$ is also a lower bound function of $E(x)$.
In the above equation, $\lambda_k$ and $w_{ij}$ ($1 \le i,j \le n$)
	are coefficients of the linear combination of $E_i$ and $\Psi^{(k)}_j (x_j)$.
$\lambda_k$ satisfies $0 \le \lambda_k < 1$ and $w_{ij}$s satisfy $w_{ij} \ge 0$ and $\sum_i w_{ij} = 1$ for $1 \le i, j \le n$.

Without loss of generality,  assume that all objective functions including the sub-objective functions $E_i(x)$ are nonnegative functions.
Then the cooperative optimization theory tells us that the lower bound function $\Psi(x)$ 
	computed by (\ref{cooperative_optimization}) can be progressively tightened, 
\[ \Psi^{(0)}(x) \le \Psi^{(1)}(x) \le \ldots \le \Psi^{(k)}(x) \le E(x) \ , \]
	when we choose the initial condition as $\Psi^{(0)}_i(x_i) = 0$, for $i=1,2,\ldots, n$.
	
The difference equations~(\ref{cooperative_optimization}) define
	the dynamics of cooperative optimization.
The original minimization problem $\min_x E(x)$ has been divided into $n$ sub-problems of minimization (see (\ref{cooperative_optimization})).
Those sub-problems can be solved in parallel in implementation.
The function $\Psi^{(k)}_i (x_i)$ is the solution at solving the $i$th sub-problem.
The objective function of the $i$th sub-problem, denoted as $\tilde{E}_i(x)$,
	is a linear combination of the original sub-objective function $E_i(x)$ and the solutions from solving other sub-problems,
	i.e.,
\[ \tilde{E}^{(k)}_i(x) = \left(1 - \lambda_k \right) E_i (x)  + \lambda_k \sum_{j} w_{ij} \Psi^{(k)}_j(x_j)	\ . \]
The cooperation among solving those sub-problems is thus achieved
	by having each sub-problem compromising its solution with the solutions of other sub-problems.
$\tilde{E}_i(x)$ is called the modified objective function for the sub-problem $i$.

The coefficient $\lambda_k$ is a parameter for controlling the cooperation at solving the sub-problems and is called the cooperation strength.
A high cooperation strength leads to strong cooperation at solving the sub-problems
	while a lower cooperation strength leads to weak cooperation.
The coefficients $w_{ij}$ control the propagation of the sub-problem solutions $\Psi_i(x_i)$ 
	in the modified objective functions $\tilde{E}_i(x)$ (details in \cite{HuangBookCCO}).
They are so called the message propagation parameters.

The function $\Psi^{(k)}_i (x_i)$ can be understood as the soft decision of assigning the variable $x_i$
	at minimizing $\tilde{E}_i(x)$.
The most preferable value for variable $x_i$ at iteration $k$ is $\arg \min_{x_i} \Psi^{(k)}_i (x_i)$.

Given any variable, say $x_i$, it may be contained in several sub-problems. 
At each iteration, 
	$x_i$ has a value in the optimal solution for each of the sub-problems.
Those values may not be the same.
If all of them are of the same value, denoted as $\tilde{x}_i$,
	we say that all the sub-problems reach a consensus assignment for variable $x_i$.
If all sub-problems reach a consensus assignment for each variable at some iteration,
	we say that a consensus solution is reached at the iteration.
Consensus solution is an important concept of cooperative optimization
	for defining global optimality conditions.
Normally, a consensus solution is the global optimum, guaranteed by theory.
The following theory~\cite{HuangBookCCO} offers one global optimality condition 
	based on the concept of consensus solution.
%
%
\begin{theorem}
\label{sufficient_condition}
Assume that the difference equations~(\ref{cooperative_optimization})
	for cooperative optimization reaches its equilibrium, 
	denoted as $(\Psi^{\infty}(x_i))$ ($1 \le i \le n$), i.e., 
	$(\Psi^{\infty}(x_i))$ is a solution to the difference equations~(\ref{cooperative_optimization}).
If a consensus solution ${\tilde x}$ is found in this case, 
	then it must be the global optimum of $E(x)$, ${\tilde x} = x^{*}$.
\end{theorem}

It has been shown in \cite{HuangBookCCO} that the cooperative optimization algorithm defined by the difference equations~(\ref{cooperative_optimization})
	has many important computational properties not possessed by conventional optimization algorithms.
Given a constant cooperation strength $\lambda_k = \lambda$ and the propagation coefficients $w_{ij}$,
	the algorithm has one and only one equilibrium.
It always converges to the unique equilibrium
	with an exponential rate regardless of initial conditions and perturbations.
Mathematical analysis also shows that when $\lambda \rightarrow 1$,
	the lower bound function computed by the difference equations~(\ref{cooperative_optimization})
	tends to have a global minimum approaching the global minimum of the original function as the iteration proceeds.
Whenever those two global minimums touch each other, the global minimum of the original function is found.

Hence, the cooperative optimization is 
	stable (unique equilibrium), 
	fast (exponential convergence rate), and robust (insensitive to initial conditions and perturbations).
Unlike conventional optimization methods, it does not struggle with local minimums
	and it knows when to stop search because of the global optimality condition.
Details about these together with the theoretical investigation of cooperative optimization
	are provided in \cite{HuangBookCCO}.

Let $\Psi^{(k)}_i (x_i) / (1- \lambda_k) \Rightarrow \Psi^{(k)}_i (x_i)$,
	the difference equations~(\ref{cooperative_optimization}) can be rewritten in a different form,
\begin{equation}
\Psi^{(k)}_i (x_i) = \min_{X_i \setminus{x_i}} E_i  +
\lambda_k \sum_{j} w_{ij} \Psi^{(k-1)}_j(x_j) \ ,
\label{cooperative_optimization2}
\end{equation}
	
With certain way of decomposing $E(x)$, 
	certain settings of the cooperation strength $\lambda_k$ and the message propagation parameters $w_{ij}$,
	the normalized min-sum algorithm can be derived 
	from the difference equations~(\ref{cooperative_optimization}) of cooperative optimization.

\section{Constructing More Powerful Cooperative Optimization Algorithms}

Describing cooperative optimization in the language of mathematics 
	enables us to make a generalization of the basic cooperative optimization algorithm 
	(see Difference Equation~(\ref{cooperative_optimization})).
One way to generalize it is to take more complicated forms of 
	the lower bound functions other than the simple one $\sum_i \Psi^{(k)}_i(x_i)$.
One of them is to break $\Psi_i(x_i)$ into several pieces as follows
\[ \Psi_i(x_i) \rightarrow \Psi_{i_1}(x_i),\Psi_{i_2}(x_i),\ldots,\Psi_{i_{N_i}}(x_i) \ . \]
Consequently, the lower bound function takes the following form
\begin{equation}
E_{-}(x_1, x_2, \ldots, x_n) = \sum_i \sum_j \Psi_{ij}(x_i) \ . 
\label{fragmented_LBF}
\end{equation}
Such a lower bound function is called the fragmented unary lower bound function.

Let $\{E_{ij}(x)\}$ be a decomposition of $E(x)$, i.e.,
\[ \sum_i \sum_j E_{ij}(x) = E(x) \ . \]
Assume that there is one sub-objective function $E_{ij}(x)$ for each $\Psi_{ij}(x_i)$.
Correspondingly, the difference equations of the cooperative optimization become
\begin{equation}
\Psi^{(k)}_{ij} (x_i) = \min_{X_{ij} \setminus{x_i}}E_{ij}  +
\lambda_k \sum_{i^{'}} \sum_{j^{'}} w_{iji^{'}j^{'}} \Psi^{(k-1)}_{i^{'} j^{'}}(x_{i^{'}}),
\label{cooperative_optimization_ex}
\end{equation}
where $X_{ij}$ is a set of variables containing those in $E_{ij}(x)$.

Our simulation has demonstrated that cooperative optimization algorithms based on the fragmented unary lower bound function
	are often more powerful at decoding LDPC codes than those based on the simple form.

\section{LDPC Decoding as Combinatorial Optimization}

LDPC codes belong to a special class of linear block codes
	whose parity check matrix $H$ has a low density of ones.
For a LDPC code over $GF(q)$, its parity check matrix $H$ has elements $h_{mn}$ defined over $GF(q)$, $h_{mn} \in GF(q)$.
Let the code word length be $N$ (the number of symbols),
then $H$ is a $M \times N$ matrix, where $M$ is the number of rows. 
Each row of $H$ introduces one parity check constraint on input data $x=(x_1, x_2, \ldots, x_N)$, i.e.,
\[ \sum^{N}_{n=1} h_{mn} x_n = 0,~~\mbox{for $m=1,2,\ldots, M$} \ . \]
Putting the $m$ constraints together, we have $H x^T = 0$.

Let function $f_n (x_n)$ be defined as
\begin{equation}
f_n (x_n) = - \ln p(x_n/y_n) \ , 
\label{unary_constraint}
\end{equation}
where $p (x_n / y_n)$ is the conditional distribution of input data symbol $n$ at value $x_n$ 
	given the output data symbol $n$ at value $y_n$.
$f_n(0) - f_n(x_n)$, which is equal to $\ln (p(x_n/y_n) / p(0/y_n))$,
	is the log-likelihood ratio (LLR) of input data symbol $n$ at value $x_n$ versus value $0$.

In those notations, the maximum likelihood decoding can be formulated as a constrained combinatorial optimization problem,
\begin{equation}
\min_{x_1, x_2, \ldots, x_N} \sum^n_{n=1} f_n (x_n) \quad \mbox{ s.t. $Hx^T = 0$} \ . 
\label{original_problem}
\end{equation}
The function to be minimized in (\ref{original_problem}) is called the objective function for decoding a LDPC code.
The decoding problem is, thus, transferred as finding the global minimum of a multi-variate objective function.

Let $X$ be the set of all variables.
Given the $m$th constraint be $H_m x^T = 0$,
	let $X_m$ be the set of variables corresponding to the non-zero elements in $H_m$, 
	i.e.,
\[ X_m \equiv \{x_n|h_{mn} \not = 0\} \ . \]
Let $f_{X_m}(X_m)$ be a function defined over $X_m$ as
\begin{equation}
f_{X_m}(X_m) = \left\{ \begin{array}{ll}
                      0, & \mbox{if $H_m x^T = 0$}; \\
                      \infty, & \mbox{ otherwise}.
                     \end{array}
             \right. 
\label{parity_check_constraint}
\end{equation}
$f_{X_m}(X_m)$ is called the constraint function representing the $m$th constraint.
Using the constraint functions, 
	the decoding problem~(\ref{original_problem}) can be reformulated as an unconstrained optimization problem
	of the following objective function,
\begin{equation}
\sum^{M}_{m=1} f_{X_m}(X_m) + \sum^N_{n=1} f_n (x_n) \ . 
\label{unconstrained_objective}
\end{equation}
Such a combinatorial optimization problem is, in general, NP-hard.

\section{Deriving the Normalized Min-Sum Algorithm}

In the following discussions,
	we differentiate unary constraints from higher order constraints in notations
	by using symbol $f_n(x_n)$ for unary constraint on variable $x_n$
	and $f_{X_m}(X_m)$ for a constraint of an order higher than one.
When constraints are mentioned, they are referred to non-unary constraints
 	and unary constraints will be explicitly declared.

Conventionally, a LDPC code is represented as a Tanner graph,
	a graphical model useful at understanding code structures and decoding algorithms.
A Tanner graph is a bipartite graph with variable nodes on one side
	and constraint nodes on the other side. 
Edges in the graph connect constraint nodes to variable nodes. 
A constraint node connects to those variable nodes that are contained in the constraint.
A variable node connects to those constraint nodes that use the variable in the constraints.
Constraint nodes are also referred to as check nodes.
During each iteration of the min-sum algorithm,
	messages are flowed from variables nodes to the check nodes first,
	then back to variable nodes from check nodes.

To follow the notation used in the literatures of coding,
	we change the symbols for indices from $i$ and $j$ to $m$ and $n$.
Let ${\cal N}(m)$
	be the set of variable nodes that are connected to the check node $m$.
Let $ {\cal M}(n)$ be the set of check nodes that are connected to the variable node $n$.
Let symbol `$\setminus$' denotes the set minus.
${\cal N}(m) \setminus n$ denotes the set of variable nodes excluding node $n$ 
	that are connected to the check node $m$.
${\cal M}(n) \setminus m$ stands for the set of check nodes excluding the check node $m$ 
	which are connected to the variable node $n$.
	
To derive the normalized min-sum algorithm as a cooperative optimization algorithm,
	we choose the fragmented unary function (see (\ref{fragmented_LBF})) 
	as the form of the lower bound function.
For each variable $n$ and then for each constraint containing the variable, 
	$ m \in {\cal M}(n)$,
	we define a component function $\Psi_{nm}(x_n)$.
The summation of all those component functions is the form of the lower bound function, 
\[ E_{-}(x_1, x_2, \ldots, x_N) = \sum^{N}_{n=1} \sum_{m \in {\cal M}(n)} \Psi_{nm}(x_n) \ . \]

Let the decomposition of the objective function $E(x)$ for a LDPC decoding problem be $\{E_{nm}(x)\}$.
Functions $E_{nm}(x)$ are sub-objective functions.
There is one sub-objective function $E_{nm}(x)$ for each $\Psi_{nm}(x_n)$.
The overall objective function is
\[ \sum_n \sum_{m \in {\cal M}(n)} E_{nm}(x) \ . \]
Different decompositions lead to different cooperative optimization algorithms.
The decomposition leading to the normalized min-sum algorithm has the following form
\begin{equation}
E_{nm}(x) = f_{X_m}(X_{m}) + \sum_{n^{'} \in {\cal N}(m) \setminus n } f_{n^{'}}(x_{n^{'}}) \ . 
\label{sub-objective_func}
\end{equation}
It contains the $m$th constraint $f_{X_m}(X_{m})$, all the unary constraints on the variables in $X_m$
	except the one on $x_n$.

Substituting (\ref{sub-objective_func}) into the difference equations~(\ref{cooperative_optimization_ex}) 
	for the fragmented unary lower bound function
	and choosing $\lambda_k w_{nmn^{'}m^{'}} = \alpha_k$ (a constant),
	we have
\begin{equation}
\Psi^{(k)}_{nm} (x_n) = \min_{X_m \setminus x_n} f_{X_m}(X_{m}) + 
	\sum_{n^{'} \in {\cal N}(m) \setminus n } Z^{(k-1)}_{n^{'}m}(x_n^{'}),
\label{cooperative_optimization3a}
\end{equation}
where 
\begin{equation}
Z^{(k-1)}_{nm}(x_n) = f_n(x_n) + \alpha_k \sum_{m^{'} \in {\cal M}(n^{'}) \setminus m^{'}} \Psi^{(k-1)}_{nm^{'}}(x_n) \ . 
\label{v_scan}
\end{equation}

For binary LDPC codes, all $x_n$s are binary variables, 
	the unary constraint $f_n(x_n)$ defined in (\ref{unary_constraint}) 
	can be rewritten as 
\begin{equation}
f_n(1) = \ln \frac{p(x_n = 0/y_n)}{p(x_n = 1/y_n)},~~\mbox{and}~~f_n(0) = 0 \  . 
\label{unary_constraint2}
\end{equation}
This change does not have any impact on the objective function~(\ref{unconstrained_objective}) of decoding
	except offsetting the objective function by a constant.
We can also offset $\Psi^{(k)}_{nm} (x_n)$ in (\ref{cooperative_optimization3a}) by $\Psi^{(k)}_{nm} (0)$.
This change also does not have any impact on the optimization results.
Putting these two changes together, the difference equations~(\ref{cooperative_optimization3a}) 
	can be simplified to
\[\Psi^{(k)}_{nm} (0) = 0, \quad \mbox{and}\]
\begin{equation}
\Psi^{(k)}_{nm} (1) = \prod_{n^{'} \in {\cal N}(m) \setminus n} \mbox{sgn}(Z^{(k-1)}_{n^{'}m}(1)) \cdot \min_{n^{'} \in {\cal N}(m) \setminus n} |Z^{(k-1)}_{n^{'}m}(1)| \ .
\label{min-sum}
\end{equation}
Eq.~(\ref{min-sum}) is the check node update rule of the normalized min-sum algorithm 
	and Eq.~(\ref{v_scan}) is the variable node update rule.
The normalized min-sum algorithm is, thus, derived from the difference equations of cooperative optimization.

Difference Equation~(\ref{cooperative_optimization3a}) defines an optimization algorithm more general than 
	the normalized min-sum algorithm.
It is suitable for any kind of variables and any kind of constraints.
The normalized min-sum algorithm defined by (\ref{min-sum}) and (\ref{v_scan}) is a special instance of the algorithm,
	applicable only to boolean variables and parity check constraints.
By understanding the underlying principle of the normalized min-sum algorithm,
	we make it possible to derive more general decoding algorithms.

\section{Objective Function of the Min-Sum Algorithm}

When we derived the normalized min-sum algorithm in the previous section,
	we did not check the objective function it tries to minimize.
From the cooperative optimization theory,
	the objective function can be obtained by summing up all sub-objective functions.
In the case of the normalized min-sum algorithm, the objective function is
\[ \sum_n \sum_{m \in {\cal M}(n)} \left( f_{X_m} (X_m) + \sum_{n^{'} \in {\cal N}(m) \setminus n} f_{n^{'}}(x_{n^{'}}) \right) \ . \]
If $f_{X_m}(X_m)$s are parity check constraints~(\ref{parity_check_constraint}),
	we can drop $f_{X_m}(X_m)$ in the above objective function and rewrite it as
\[ \sum_n \sum_{m \in {\cal M}(n)} \sum_{n^{'} \in {\cal N}(m) \setminus n} f_{n^{'}}(x_{n^{'}}), \quad \mbox{s.t. $Hx^T = 0$} \ . \]
This objective function is, in general, not proportional to 
	$\sum_n f_n (x_n)$
except of a few cases, e.g., regular LDPC codes.
A regular LDPC code has the property that all variable nodes have the same degree and 
	all the check nodes also have the same degree.
Otherwise, it is called an irregular LDPC code.

Let $d_v(n)$ be the degree of the variable node $n$.
Let $d_c(m)$ be the degree of the check node $m$.
Using these notations, the objective function of the normalized min-sum algorithm becomes
\[ \sum_n A_n f_n (x_n), \quad \mbox{where}\]
\begin{equation}
A_n = \sum_{m \in {\cal M}(n)} (d_c(m) - 1), \quad \mbox{for $n=1, 2, \ldots, N$} \ .
\label{balance_a}
\end{equation}
The above objective function is proportional to the desired objective function $\sum_n f_n (x_n)$
	if and only if all $A_n$s have the same value for all $n$s.

Assume that all constraints in a LDPC code have the same degree, $d_c$, then $A_n$ in Formula~(\ref{balance_a}) becomes
\[ (d_c - 1) d_v (n), \quad \mbox {for $n =1,2, \ldots, N$} \ . \]
Hence the weight for each variable in the objective function is proportional to the degree of the variable.
The objective function of the normalized min-sum algorithm is biased in this case.
It weights bit variables of higher degrees more than the variables of lower degrees.
That explains why we have experienced less decoding errors for bit variables of higher degrees 
	in applying the normalized min-sum algorithm at decoding LDPC codes in practice.

For a regular LDPC code, $d_c(m) = d_c$ (a constant) and 
	$d_v(n) = d_v$ (a constant).
In this case, $A_n = (d_c - 1) d_v$ for all $n$,
	and the objective function of the min-sum algorithm is proportional to $\sum_n f_n(x_n)$,
	which is the desired objective function for the maximum likelihood decoding of the code.

\section{General Framework for Constructing Min-Sum Algorithms}

Viewing the normalized min-sum algorithm as an instance of cooperative optimization
	lays out several new ways to design new decoding algorithms.
One way is to select different forms of the lower bound functions 
	besides	the simple form and the fragmented form offered in this paper.
Another way is to explore different decompositions 
	of the objective function associated with a decoding problem.
The third way is to use different settings of the cooperation strength $\lambda_k$ 
	and message propagation parameters $w_{ij}$.
The decomposition presented in this paper is simple and direct.
It is only one of many possible ways of decomposing an objective function.
Correspondingly, representing an objective function using a Tanner graph,
	different decompositions of the objective function 
	correspond to different ways of decomposing the Tanner graph into sub-graphs.
Each LPDC code represented as a Tanner graph may also have its own unique graphical structure.
Special decomposition can also be explored to maximize the power of cooperative optimization at decoding the code.

\section{Conclusions}

This paper has shown that the normalized min-sum algorithm is
	the simplification of a cooperative optimization algorithm
	for the special case of boolean variables and parity check constraints. 
A generalized min-sum algorithm for decoding LDPC codes over $GF(q)$ , for any $q \ge 2$,
	and any form of check constraints are also offered in this paper~(see Eq.~\ref{cooperative_optimization3a}).
	
The normalized min-sum algorithm can be understood as
	finding the global minimum via a lower bounding technique.
To decode a LDPC code, it deploys a function of a simple form
	which is a lower bound function to the objective function associated with the decoding problem.
To find the global optimum of the original objective function,
	it progressively tightens the lower bound function
	until the global minimum of the lower bound function reaches the global minimum of the original objective function.
	
A consensus solution found by a cooperative optimization algorithm
	is in general the global optimum.
It offers a general criterion for the normalized min-sum algorithm 
	to identify the optimal codeword and to terminate its search process.
The criterion is that each variable of the problem has the same best assignment 
	in minimizing the modified sub-objective functions defined by the cooperative optimization (see Eq.~\ref{cooperative_optimization3a}).

It has also been shown that the normalized min-sum algorithm has a biased objective function
	with higher weights for those variables of higher degrees.
The objective function becomes the desired, unbiased objective function	when a LDPC code is regular.

\nocite{HuangISIT05,Yedidia05,Wainwright05,Richardson:DOCAILDPCC}
\bibliographystyle{../bib/IEEEtran}

\begin{thebibliography}{10}
\providecommand{\url}[1]{#1}
\def\UrlFont{\rmfamily}
\providecommand{\newblock}{\relax}
\providecommand{\bibinfo}[2]{#2}
\providecommand\BIBentrySTDinterwordspacing{\spaceskip=0pt\relax}
\providecommand\BIBentryALTinterwordstretchfactor{4}
\providecommand\BIBentryALTinterwordspacing{\spaceskip=\fontdimen2\font plus
\BIBentryALTinterwordstretchfactor\fontdimen3\font minus
  \fontdimen4\font\relax}
\providecommand\BIBforeignlanguage[2]{{%
\expandafter\ifx\csname l@#1\endcsname\relax
\typeout{** WARNING: IEEEtran.bst: No hyphenation pattern has been}%
\typeout{** loaded for the language `#1'. Using the pattern for}%
\typeout{** the default language instead.}%
\else
\language=\csname l@#1\endcsname
\fi
#2}}

\bibitem{Jinghu2002}
J.~Chen and M.~Fossorier, ``Density evolution of two improved {BP}-based
  algorithms for {LDPC} decoding,'' \emph{IEEE Communications Letters}, vol.~6,
  pp. 208--210, 2002.

\bibitem{JinghuChenThesis}
J.~Chen, ``Reduced complexity decoding algorithms for low-density parity check
  codes and turbo codes,'' Ph.D. dissertation, University of Hawaii, Dept. of
  Electrical Engineering, 2003.

\bibitem{Kschischang01}
F.~R. Kschischang, B.~J. Frey, and H.~andrea Loeliger, ``Factor graphs and the
  sum-product algorithm,'' \emph{IEEE Transactions on Information Theory},
  vol.~47, no.~2, pp. 498--519, February 2001.

\bibitem{Aji97}
S.~Aji and R.~McEliece, ``The generalized distributive law,'' \emph{IEEE
  Transactions on Information Theory}, vol.~46, pp. 325--343, March 2000.

\bibitem{Berrou93}
C.~Berrou, A.~Glavieux, and P.~Thitimajshima, ``Near shannon limit
  error-correcting coding and decoding: turbo codes,'' in \emph{Proceedings of
  the 1993 IEEE International Conference on Communication}, 1993, pp.
  1064--1070.

\bibitem{Gallager:LDPCC:thesis}
R.~G. Gallager, ``Low-density parity-check codes,'' Ph.D. dissertation,
  Department of Electrical Engineering, M.I.T., Cambridge, Mass., July 1963.

\bibitem{MacKay:GCBOVSM}
D.~J.~C. MacKay and R.~M. Neal, ``Good codes based on very sparse matrices,''
  in \emph{Cryptography and Coding, 5th IMA Conference}, December 1995.

\bibitem{Wiberg:thesis}
N.~Wiberg, ``Codes and decoding on general graphs,'' Ph.D. dissertation,
  Department of Electrical Engineering, Linkoping University, Linkoping,
  Sweden, 1996.

\bibitem{forney73}
J.~G.~D.~Forney, ``The {Viterbi} algorithm,'' \emph{Proc. IEEE}, vol.~61, pp.
  268--78, Mar. 1973.

\bibitem{Pearl88}
J.~Pearl, \emph{Probabilistic Reasoning in Intelligent Systems: Networks of
  Plausible Inference.}\hskip 1em plus 0.5em minus 0.4em\relax Morgan Kaufmann,
  1988.

\bibitem{Fossorier99}
M.~Fossorier, M.~Mihaljevic, and H.~Imai, ``Reduced complexity iterative
  decoding of low density parity check codes based on belief propagation,''
  \emph{IEEE Transactions on Communications}, vol.~47, pp. 673--680, May 1999.

\bibitem{Hagenauer96}
J.~Hagenauer, E.~Offer, and L.~Papke, ``Iterative decoding of binary block and
  convolutional codes,'' \emph{{IEEE} Transactions on Information Theory},
  vol.~42, pp. 1064--1070, 1996.

\bibitem{Shannon48}
C.~E. Shannon, ``A mathematical theory communication,'' \emph{Bell Sys. Tech.
  J.}, vol.~27, pp. 379--423,623--656, 1948.

\bibitem{HuangBookCCO}
X.~Huang, ``Cooperative optimization for solving large scale combinatorial
  problems,'' in \emph{Theory and Algorithms for Cooperative Systems}, ser.
  Series on Computers and Operations Research.\hskip 1em plus 0.5em minus
  0.4em\relax World Scientific, 2004, pp. 117--156.

\bibitem{HuangISIT05}
------, ``Near perfect decoding of {LDPC} codes,'' in \emph{Proceedings of IEEE
  International Symposium on Information Theory (ISIT)}, 2005, pp. 302--306.

\bibitem{Yedidia05}
J.~Yedidia, W.~Freeman, and Y.~Weiss, ``Constructing free-energy approximations
  and generalized belief propagation algorithms,'' \emph{IEEE Transactions on
  Information Theory}, vol.~51, no.~7, pp. 2282--2312, July 2005.

\bibitem{Wainwright05}
M.~Wainwright, T.~Jaakkola, and A.~Willsky, ``A new class of upper bounds on
  the log partion function,'' \emph{IEEE Transactions on Information Theory},
  vol.~51, no.~7, pp. 2313--2335, July 2005.

\bibitem{Richardson:DOCAILDPCC}
T.~J. Richardson, M.~A. Shokrollahi, and R.~L. Urbanke, ``Design of
  capacity-approaching irregular low-density parity-check codes,'' \emph{{IEEE}
  Transactions on Information Theory}, vol.~47, no.~2, pp. 619--637, February
  2001.

\end{thebibliography}

\end{document}